\documentclass[twocolumn,showpacs,preprintnumbers,amsmath,amssymb,groupedaddress]{revtex4-1}
\usepackage[english]{babel}
\usepackage[combine]{ucs}
\usepackage[utf8x]{inputenc}
\usepackage[dvips]{graphicx}
\usepackage{amssymb}

\begin{document}

\author{Jani Kotakoski}
\email[Corresponding author. Email: ]{jani.kotakoski@iki.fi}
\affiliation{University of Vienna, Department of
  Physics, Boltzmanngasse 5, 1090 Wien, Austria}
\author{Jannik C. Meyer}
\affiliation{University of Vienna, Department of
  Physics, Boltzmanngasse 5, 1090 Wien, Austria}

\title{Mechanical properties of polycrystalline graphene based on a
  realistic atomistic model}

\begin{abstract}
  Graphene can at present be grown at large quantities only by the
  chemical vapor deposition method, which produces polycrystalline
  samples. Here, we describe a method for constructing realistic
  polycrystalline graphene samples for atomistic simulations, and
  apply it for studying their mechanical properties. We show that
  cracks initiate at points where grain boundaries meet and then
  propagate through grains predominantly in zigzag or armchair
  directions, in agreement with recent experimental work. Contrary to
  earlier theoretical predictions, we observe normally distributed
  intrinsic strength ($\sim 50$\% of that of the mono-crystalline
  graphene) and failure strain which do not depend on the
  misorientation angles between the grains. Extrapolating for grain
  sizes above 15~nm results in a failure strain of $\sim 0.09$ and a
  Young's modulus of $\sim 600$~GPa. The decreased strength can be
  adequately explained with a conventional continuum model when the
  grain boundary meeting points are identified as Griffith cracks.
\end{abstract}

\maketitle

Grain boundaries define the electronic and mechanical properties of
polycrystalline materials. Since chemical vapor deposition (CVD) is
currently the only way for producing industry-scale graphene
membranes, and leads to polycrystalline samples, study of grain
boundaries in graphene has become of fundamental importance during the
recent years. In a two-dimensional material, such as graphene, the
boundaries also have a critical contribution to the chemical
reactivity. Because of this, although atomic scale imaging can in
principle reveal their exact structure, the boundaries tend to be
covered by adsorbates with only short segments available for direct
imaging. Nevertheless,
experiments~\cite{Cervenka2009a,Huang2011,An2011,Kim2011,Kurasch2012,Tapaszto2012}
have revealed meandering serpent-like boundaries which are typically
formed from pentagon-heptagon--pairs in the parts not covered by the
adsorbates.

Mechanical properties of graphene sheets have been a topic of intense research
already for two decades in the context of carbon nanotubes (see
Ref.~\cite{Yakobson2001} for a topical review). More recently, in
2007~\cite{Liu2007}, Liu and co-workers utilized {\it ab initio} calculations
to study the elastic moduli and fracture characteristic of graphene. Young's
modulus was found to be 1.05~TPa, and failure strain, depending on the pulling
direction, 0.194--0.266. Intrinsic strength was estimated to be 110--121~GPa,
similarly depending on the pulling direction. The role of pre-existing defects
on these properties was also studied~\cite{Khare2007}. It was noticed that
their effect does not depend on the exact atomic structure of the defects but
rather on their size. The authors also showed that the intrinsic strength of
graphene with crack-like defects can be described with a continuum model using
the Griffith formula for defect sizes down to 10~{\AA}. Soon after this, Frank
{\it et al.}  used a tip of an atomic force microscope to obtain a Young's
modulus of 0.5~TPa for suspended stacks of graphene sheets~\cite{Frank2007}. A
year later, Lee and co-workers reported on several mechanical properties of
graphene using a similar technique~\cite{Lee2008}, establishing graphene as
the strongest material ever measured. They reported an intrinsic strength of
42~N/m (corresponding to 130~GPa assuming graphene thickness to be the
inter-layer distance in graphite, i.e., 0.335~nm) occurring at 0.25 strain.
Young's modulus was estimated to be 1~TPa, in a good agreement with
theory~\cite{Liu2007}. In 2009, Xiao and others reported on their theoretical
work~\cite{Xiao2009}, in which they obtained failure strain of ca. 0.10 for
graphene sheets with Stone-Wales defects (one rotated bond) with an intrinsic
strength very close to that of the pristine structure (the difference was
larger for small-diameter nanotubes). This result would be consistent with the
continuum model~\cite{Khare2007} assuming the defect corresponds to a crack
with a size below 5~{\AA}, which seems reasonable for this defect.

These early works concentrated on either mechanically exfoliated
pristine graphene or mono-crystalline graphene with point defects. The
first experimental studies on mechanical properties of polycrystalline
graphene samples were carried out only last year
(2011)~\cite{Huang2011,Ruiz-Vargas2011,Kim2011a}. The experimental
results can be summarized as follows: The intrinsic strength for
polycrystalline samples is somewhat above one third of that for
mono-crystalline graphene
(ca.~35~GPa)~\cite{Huang2011,Ruiz-Vargas2011} and cracks propagate
through the bulk of the grains~\cite{Kim2011a} mostly along zigzag and
armchair directions, not along grain boundaries as could be
intuitively expected.

In the meanwhile, also theoretical work on energetics and other non-elastic
properties~\cite{Malola2010,Yazyev2010,Carlsson2011,Kurasch2012} of
grain boundaries as well as on their mechanical
response~\cite{Grantab2010,Ruiz-Vargas2011} has been carried
out. Total energy calculations~\cite{Yazyev2010,Carlsson2011} have
established that the idealized low-energy configuration of grain
boundaries is a linear tilt boundary consisting of a repeating set of
pentagon-heptagon--pairs which act as dislocation cores in a lattice
otherwise constructed of hexagonal carbon rings. Intrinsic strength of
graphene sheets with infinitely long such boundaries has been estimated to be
50--100~GPa with failure strains in the range of
0.07--0.15~\cite{Grantab2010}, depending on the misorientation angle between
the adjacent grains. Since a higher misorientation angle yields a higher
density of dislocation cores at the linear tilt boundary, but also to higher
intrinsic strength of the model structures, the authors noted that their
results disagree with the fracture mechanics model, assuming the heptagons of
the dislocation cores correspond to Griffith cracks, which would predict
graphene sheets to become weaker with an increasing defect density. If the
grain boundaries themselves are indeed the weakest point in the lattice, this
can be argued to be a reasonable comparison since also Stone-Wales defect
consists of pentagons and heptagons and it has been shown to weaken
graphene~\cite{Xiao2009}. However, despite similarities between the
theoretical models~\cite{Yazyev2010,Carlsson2011} and short segments of the
actually observed non-decorated
boundaries~\cite{Cervenka2009a,Huang2011,An2011,Kim2011,Kurasch2012}, it
remains unclear whether such infinitely long linear arrangements of
dislocation cores can serve as a realistic model for studying mechanical
properties of polycrystalline graphene.

Theoretical calculations presented along the experimental work in
Ref.~\cite{Ruiz-Vargas2011} assumed that voids would exist in polycrystalline
graphene samples and that they could explain the apparent discrepancy between
the experimental results of $\sim 35$~GPa and the theoretical estimates of
50--100~GPa. However, it is questionable how well this model corresponds to
actual polycrystalline graphene samples. Moreover, the inherent difficulties
in assessing the mechanical properties of a membrane suspended on a hole by
applying force with a tip of a microscope necessitate theoretical confirmation
with a realistic model system.

Here, we describe an automated method for creating polycrystalline
graphene structures with realistic misorientation angle and carbon
ring size statistics as well as serpent-like boundaries similar to
those observed experimentally. Using atomistic simulations, we then
subject our samples to a study of their mechanical properties. We show
that close to the failure strain, cracks appear typically at the
points where grain boundaries meet, and in agreement with the recent
experimental studies, then propagate through grains predominantly in
zigzag or armchair directions.  Contrary to earlier theoretical
predictions~\cite{Grantab2010}, neither intrinsic strength nor failure
strain of our samples depend on the misorientation angle between the
grains, but are normally distributed similar to recent experimental
studies,~\cite{Huang2011,Ruiz-Vargas2011} where intrinsic strength of
$\sim 35$~GPa was reported. We obtain a slightly higher value (46~GPa)
which is still in a reasonably good agreement with the experimental
one. At the large grain size limit ($\gg 15$~nm) the failure strain is
about 0.09 and Young's modulus is close to 600~GPa. The formation of cracks
at the meeting points of grain boundaries, completely neglected in the
previous theoretical studies, resolves the discrepancy between the
experiments and the theoretical results and shows that the Griffith
model can after all be used to describe the mechanical properties of
polycrystalline graphene samples when a realistic atomistic model is
used.

Without pre-patterned seeds for growth, CVD growth of graphene is
initiated at several nucleation sites simultaneously. On a substrate
such as Cu, which doesn't offer epitaxiality, the lattice orientations
of the growing grains are random~\cite{Yu2011}. To mimic such growth,
we first wrote a computer code which creates a pre-selected number of
randomly placed nucleation sites on a plane with pre-defined
dimensions. For each such nucleation site ($i$) a random orientation
$\theta_i$ is selected for the lattice. In order to obtain
approximately uniform size distribution for the grains, the sites are
required to be at least 5.0~{\AA} apart from each other (5~{\AA} was
selected arbitrarily). Next, we carry out an iterative process in
which any of the missing neighbors of the already inserted atoms can
appear with the same probability. When two grains approach, we use the
following condition for deciding whether a lattice site is available
for another atom: if $d<1.0$~{\AA} or $N>3$ ($d$ is the distance
between the lattice site and the closest existing atom, $N$ is the
number of atoms created closer than $a-1.0$~{\AA} to the present site,
where $a$ is the length of the graphene lattice vector) the site is
not free and will thus be disregarded for further growth. Upon
testing, this condition was found to minimize the dangling bond
density at the boundaries.

\begin{figure}[h]
\includegraphics[width=0.48\textwidth]{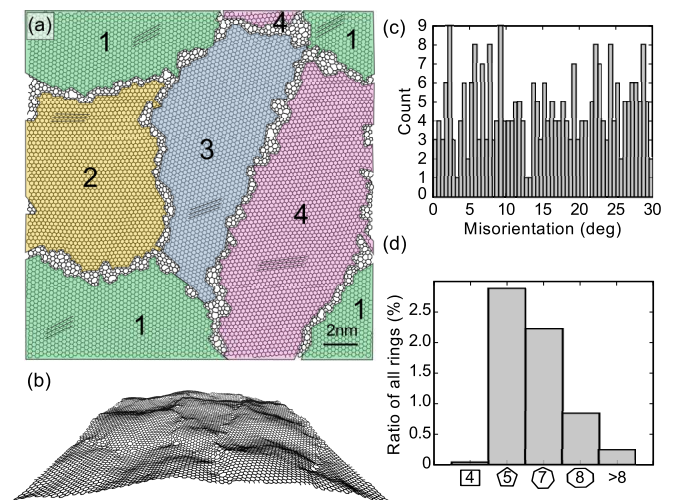}
\caption{(Color online) Model structures for polycrystalline
  graphene. (a) Top and (b) side view of a periodic 20~nm $\times$
  20~nm graphene sheet with four grains, as marked by the numbered
  shaded areas. The lines indicate orientations of the graphene
	lattice within each grain. Note that all the presented mechanical studies
	have been carried out for bicrystalline samples. (c) Distribution of
	misorientation angles for the bicrystalline sample structures used in this
	study. (d) Relative probabilities for non-hexagonal carbon rings in the same
	structures.}
\label{px::gbs}
\end{figure}

To equilibrate thus created polycrystalline samples, we first
annealed them at 3000~K for 50~ps after which the system was quenched
during a 10~ps run allowing the lattice to obtain its equilibrium size
(pressure driven to zero). At this point the lattice appears somewhat
crumpled even after pressure relaxation since we did not restrict
relaxation in the out-of-plane direction. All simulations were carried
out with the classical molecular dynamics (MD) code
{\scshape{parcas}}~\cite{Nordlund1995,Nordlund1998a,Ghaly1999} with a
reactive bond order potential developed by Brenner {\it et
  al}.~\cite{Brenner2002}.  Due to the large number of atoms in the
structures (up to almost 10000) and large number of structures (385 in
total), this is the only feasible method for carrying out the
simulations. A similar simulation setup has been used in earlier
theoretical studies of mechanical properties of
graphene~\cite{Grantab2010}, where a good agreement with {\it ab
  initio} methods has been noticed. Temperature and pressure control
were handled using the Berendsen method~\cite{Berendsen1984}. The
equilibration procedure leads to grain boundary structures similar to
the 20~nm $\times$ 20~nm model presented in
Fig.~\ref{px::gbs}a,b.

After establishing a method for creating model structures for
polycrystalline graphene, we applied it for creating 385 bicrystalline
structures with grain sizes between $\sim 3$--16~nm. Two randomly placed seeds
were used for each structure to obtain exactly one misorientation angle
($\theta$) per structure. The resulting distribution of $\theta$ is presented
in Fig.~\ref{px::gbs}c, where $\theta = \theta' = |\theta_1-\theta_2|$ if
$\theta' \leq 30^\circ$ and $\theta = 60^\circ - \theta'$ otherwise (for
graphene any $\theta_i \in [0^\circ,60^\circ]$). As expected for two randomly
selected orientations, the distribution is uniform with fluctuations resulting
only from the finite sample size. In Fig.~\ref{px::gbs}d, we plot the relative
occurrence of carbon rings other than hexagons within the created structures.
The combined likelihood for tetragons and pentagons is similar to that of
heptagons and octagons indicating mostly saturated bonds at the boundaries.
The significantly lower probability for rings with more than eight atoms is a
sign of an existing but small local density deficit at the boundaries.
Overall, the ring statistics seem reasonable. We noticed only very rarely if
ever four-coordinated atoms which would indicate problems with the interaction
model (such coordination is never observed in
$sp^2$-bonded graphene even when it is heavily amorphized under an
electron beam~\cite{Kotakoski2011,Meyer2012}). However, almost all
structures contain a few under-coordinated atoms which could serve as
reactive sites for covalently bonding adsorbates on the grain
boundaries. Thus, the experimentally observed high coverage of grain
boundaries gives further credibility for our model structures.

\begin{figure}[h]
\includegraphics[width=0.48\textwidth]{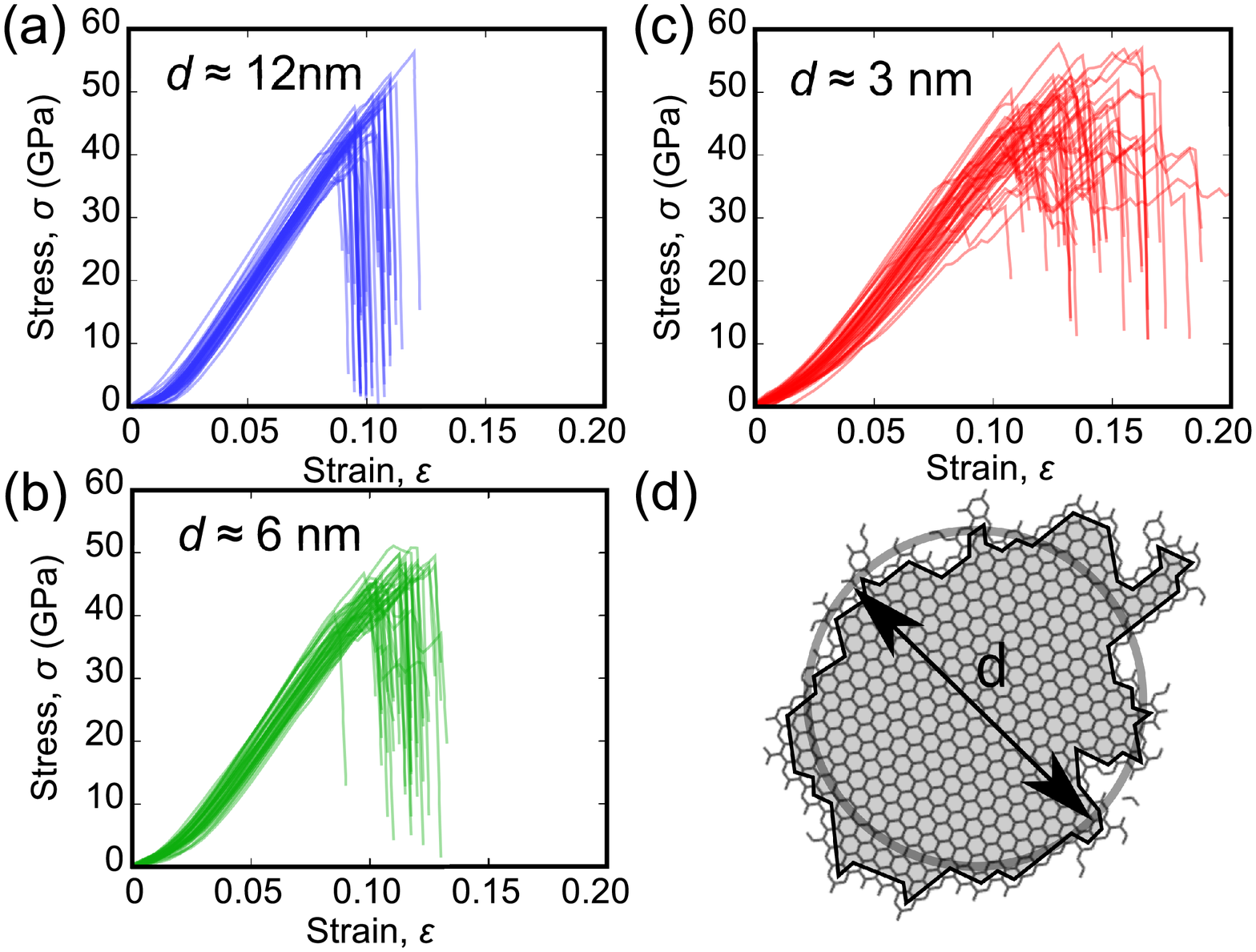}
\caption{(Color online) Stress-strain curves for polycrystalline
  graphene samples as plotted for grain sizes of (a) $d \approx
  12$~nm, (b) $d\approx 6$~nm and (c) $d \approx 3$~nm. The size of
  the grains is determined as an average diameter for a grain assumed
  to be circular as indicted in panel (d).}
\label{px::stressstrain}
\end{figure}

Next, we subjected the created sample structures to extensive tensile
testing.  The simulations were carried out at 300~K as follows: we
applied uni-axial strain in a step-by-step fashion always
equilibrating the structure for 5~ps before increasing the strain
(much slower pulling was also tested with no apparent change in the
results). We employed periodic boundary conditions for all simulations in the
in-plane ($x$ and $y$) directions. For this part of the simulations we
modified the cutoff of the interaction model to remove the unphysical
softening at longer inter-atomic distances (above $1.92$~{\AA}), which is
crucial for many MD simulations, but in the present case only affects at high
strains by leading to nonphysical features in the stress-strain curve. For
inter-atomic distances below $1.92$~{\AA}, the interaction model remained
unchanged. Modifying the cutoff has also before been noticed to be required
for properly addressing the mechanical properties of graphene close to the
fracture strain~\cite{Grantab2010}. The continuity of atomic trajectories and
conservation of energy were monitored during the simulations to avoid any
problems resulting from this modification. We also carried out test
simulations to check that the differences caused to the stress-strain
curves -- either in pristine graphene or our bicrystalline samples -- were
limited to the unphysical features near fracture.

The stress during deformation was calculated from the Virial expression as
explained in Ref.~\cite{Grantab2010} assuming a thickness of 0.335~nm for the
graphene membrane. The resulting stress-strain curves for grains with three
different average sizes are presented in Fig.~\ref{px::stressstrain}. For
pristine graphene, our method yields an intrinsic strength of 90--100~GPa at a
failure strain of 0.15--0.20, depending on the pulling angle, in a good
agreement with the {\it ab initio} results~\cite{Liu2007}. As can be seen in
Fig.~\ref{px::stressstrain}, the presence of grain boundaries leads to
approximately a 50\% reduction of the strength of the material
independent of the grain size (intrinsic strength corresponds to the
maximum stress before the failure). For grain sizes above $\sim
12$~nm, the average fracture strain is close to 0.10
(Fig.~\ref{px::stressstrain}a), whereas for the smaller ones it
increases gradually (see Figs.~\ref{px::stressstrain}b,c) up to
ca. 0.15 for the smallest reasonable grain sizes ($\sim 3$~nm). This
is because the grain boundaries are more flexible than the bulk of the
grains, and their role is pronounced at small grain sizes allowing
higher overall strains. We point out that the Poisson effect has not been
taken into account in the presented data. However, we checked whether it would
affect the results by carrying out a subset of the simulations also without
periodic boundaries in the $y$-direction. The observed deviations were within
the uncertainties stemming from the finite sample size (that is, those seen in
Fig.~\ref{px::stressstrain}).

\begin{figure}[h]
\includegraphics[width=0.48\textwidth]{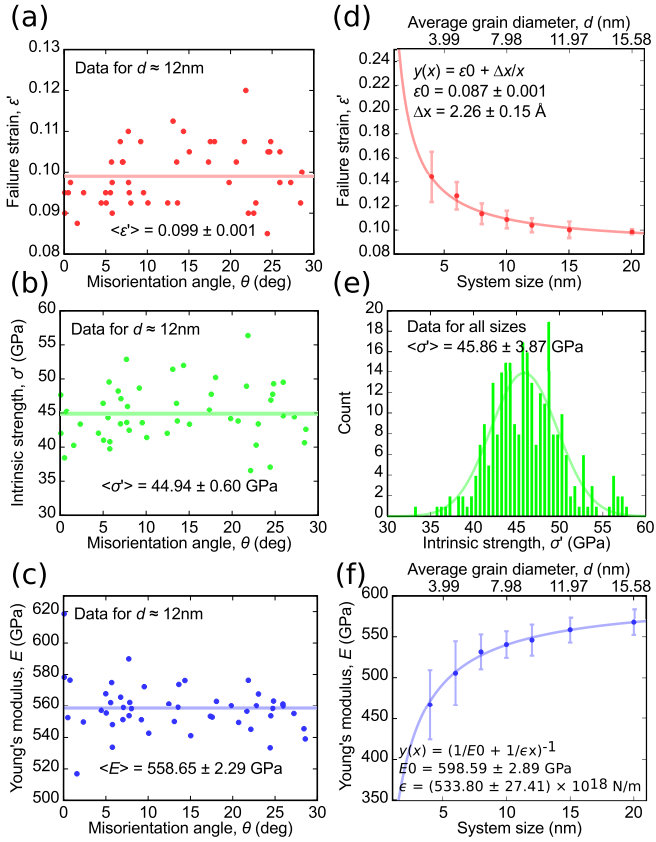}
\caption{(Color online) Mechanical properties of the sample structures
  as a function of the misorientation between the two grains
  $\theta$ and the grain size $d$ (or system size). (a) Failure
  strain, (b) intrinsic strength and (c) Young's modulus as a function
  of $\theta$ for $d\approx 12$~nm grains. (d) Failure strain as a
  function of $d$. (e) Distribution of the intrinsic strengths for all
  $d$. (f) Young's modulus as a function of $d$.}
\label{px::missangle}
\end{figure}

To better understand how the different measured properties depend on
the grain size ($d$) and misorientation between the grains ($\theta$),
we plot in Fig.~\ref{px::missangle} the failure strain, intrinsic
strength and Young's modulus as functions of $\theta$ and $d$. For the
data plotted as a function of $\theta$ (Fig.~\ref{px::missangle} a-c),
we used only one grain size ($d \approx 12$~nm) to ease the
interpretation of the results. What can be readily observed is that
none of the calculated properties depend on $\theta$. Instead, they
are normally distributed over an average value, in a stark contrast to
the earlier theoretical prediction based on infinitely long linear
grain boundary structures~\cite{Grantab2010}. However, when the
failure strain is plotted as a function of $d$
(Fig.~\ref{px::missangle}d), a clear size-dependency emerges, as was
qualitatively described above. For intrinsic strength
(Fig.~\ref{px::missangle}e) we observe no $d$-dependency at
all. Instead, the data is normally distributed for all $d$ around a
value of $\sim 46$~GPa. While Young's modulus is defined as the change
in stress divided by the change in strain for the linear part of the
stress-strain curve (Fig.~\ref{px::stressstrain}), its $d$-dependency
is determined by that of the strain (Fig.~\ref{px::missangle}f). For
the strain, we can describe the different contributions of the bulk of
the grain and the grain boundary with a constant describing the
large-grain-size-limit $\epsilon_0$ and term inversely proportional to
the grain size ($\propto d^{-1}$) to obtain $\epsilon = \epsilon_0 +
const./d$. Fitting this equation to the data for the failure strain
yields a very good agreement as can be seen in
Fig.~\ref{px::missangle}d. Through the fit we obtain failure strain
for large grains of $\sim 0.09$. A similar fit for the Young's modulus
(Fig.~\ref{px::missangle}f) gives a value of $\sim 600$~GPa similarly
for large grains.

\begin{figure*}[ht!]
\includegraphics[width=0.80\textwidth]{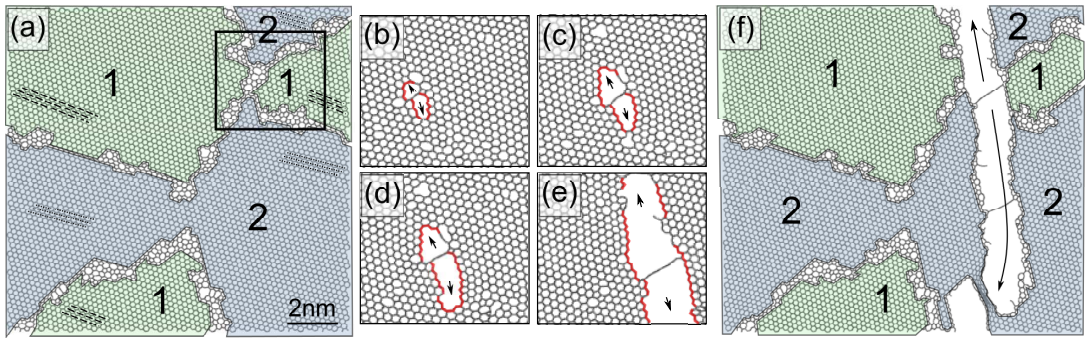}
\caption{(Color online) An example case of crack formation and
  propagation. (a) The structure with no strain with different grains
  (1 and 2) marked with different colors. The lines indicate the
  orientations of the grains. The square shows the area where the grain
  boundaries meet and the crack will appear. (b-e) Snapshots of the
  area immediately around the crack during straining showing how the
  crack penetrates along the zigzag axis of the bulk of the grain. (f)
  The structure after the crack has penetrated through the grains.}
\label{px::crack}
\end{figure*}

To further understand why the breaking stress is $d$-independent, we
visually analysed the evolution of the atomic structure of our samples
upon fracture. An example is presented in Fig.~\ref{px::crack}. What
we noticed is that the crack formation often occurs at the points where
the grain boundaries meet (marked with a square in
Fig.~\ref{px::crack}a). After the crack is formed, however, it
propagates typically along the armchair or zigzag lattice directions
in the bulk of the neighboring grains, similar to what has been
recently suggested based on experimental
observations~\cite{Kim2011}. While the atomic structure of the
boundaries, and that of their meeting points, are independent on the
grain size, the fracture properties must also be grain
size-independent, which is exactly what we observe in our
results. Moreover, the characteristic size of these meeting points in
our structures (as can be seen in Fig.~\ref{px::gbs}a and
Fig.~\ref{px::crack}a) is $\sim 2$~nm, which can be compared to the
Griffith model data from Ref.~\cite{Khare2007}. The data would
indicate that such a crack size would result in roughly 50\% reduction
in the strength of the material, in agreement with our data.

As a conclusion, we established a method for creating realistic
polycrystalline graphene samples for atomistic simulations. We applied
this method for creating a representative set of samples for
mechanical testing, and showed that the presence of grain boundaries
reduces the strength of graphene by about 50\% (down to $\sim
46$~GPa), in a reasonable agreement with
experiments~\cite{Huang2011,Ruiz-Vargas2011}. However, we observed no
misorientation-dependency on any of the mechanical properties of the
created samples which was recently suggested based on a theoretical
study on graphene structures with infinitely long linear grain
boundaries~\cite{Grantab2010}. Furthermore, we showed that crack
formation occurs at points at which the boundaries meet and that the
cracks propagate through the bulk of the neighboring grains typically
along armchair and zigzag directions, similar to recent experimental
findings~\cite{Kim2011}. The failure strain for polycrystalline
graphene with grain sizes $\gg 15$~nm was found to be $\sim 0.09$ with
a corresponding Young's modulus of $\sim 600$\ GPa. Overall our
results show that the mechanical properties of polycrystalline
graphene can be reasonably well described using the continuum model
if the grain boundary meeting points are identified as the Griffith cracks in
this material.


\end{document}